\documentclass[aip,jcp,numerical,reprint,floatfix]{revtex4-1}
\pdfoutput=1

\usepackage{graphics} 
\usepackage{amsmath} 
\usepackage{amssymb}  
\usepackage{bm}
\usepackage{enumitem}
\usepackage{algorithmic}
\usepackage{algorithm}
\usepackage{tikz}
\usepackage{hyphenat}
\usepackage{color}
\usepackage{ulem}

\newcommand*\rfrac[2]{{}^{#1}\!/_{#2}}
\begin{document}
\title{
A preconditioning scheme for Minimum Energy Path finding methods
}
\author{Stela Makri}
\affiliation{Warwick Centre for Predictive Modelling,
						School of Engineering, University of Warwick,
						CV4 7AL, Coventry, UK}
\author{Christoph Ortner}
\affiliation{Mathematics Institute,
        University of Warwick, CV4 7AL, Coventry, UK}
\author{James R. Kermode}
\affiliation{Warwick Centre for Predictive Modelling,
						School of Engineering, University of Warwick,
						CV4 7AL, Coventry, UK}

\begin{abstract}
  Popular methods for identifying transition paths between
	energy minima,
	such as the nudged elastic band and string methods, typically do not incorporate
  potential energy curvature information, leading to slow relaxation to
  the minimum energy path for typical potential energy surfaces encountered in molecular simulation.
	We propose a preconditioning
  scheme which, combined with a new adaptive timestep selection algorithm,
  substantially reduces the computational cost of transition path finding algorithms.
	We demonstrate the improved performance of our approach in a range of
  examples including vacancy and dislocation migration modelled with both interatomic potentials and density functional theory.
\end{abstract}

\maketitle

\section{Introduction}

In computational chemistry, structural biology, materials science and
 engineering, the time taken for processes is often dominated by
 transitions between energy minima in a potential energy landscape.
 The computational evaluation of the Minimum Energy Path (MEP)
 of the transition is a familiar technique used to find the energy barrier
 $\Delta E$ of such a transition~\cite{Voter2002}.
 The objective is the evaluation of the transition rate to leading order which
 is given by $\nu\sim\nu_0 \exp{\left(\rfrac{-\Delta E}{k_\mathrm{B}T}\right)}$
 ~\cite{1935JChPh...3..107E,reactionRateTheoryReview}, where the attempt rate $\nu_0$
 may be estimated using Eyring's heuristic derivation~\cite{1935JChPh...3..107E},
 or approximated with Harmonic Transition State Theory~\cite{VINEYARD1957121},
 $k_\mathrm{B}$ is the Boltzmann constant and $T$ is the temperature of the system.
 Knowing the transition rate enables the simulation of the transition on
 the mesoscale using, for example, the
 kinetic Monte Carlo method~\cite{10.1007/978-1-4020-5295-8_1}.

We restrict our focus to `double ended' cases where both energy minima are known. The most
 notable techniques in this case are the string method
 ~\cite{PhysRevB.66.052301,simplifiedString,Cameron2011} and the Nudged Elastic
 Band (NEB) method~\cite{1998cqdc.conf..385J,doi:10.1063/1.1323224}. Both
 methods find the MEP by iteratively relaxing a
 discretised path, of $N$ images, until convergence to an approximate MEP is achieved.
 Typically, the path is evolved in the energy landscape via a steepest descent-like optimisation
 technique, which may converge slowly when the potential is ill-conditioned,
 that is, the Hessian matrix of the potential
 along the path has a large condition number~\cite{opacb1120179}. Such
 a situation arises, for example, in large computational domains or if
 bonds with significant stiffness variations are present.
 Preconditioning is commonly used in linear algebra and optimisation to
 effectively reduce the
 condition number and thus improve the rate of convergence of an iterative
 scheme~\cite{opacb1120179}.

It has been shown for example in Refs.~\onlinecite{2016JChPh144p4109P,Mones2018,LINDH1995423}
how to construct and
invert effective preconditioners for the potential energy landscape of materials
and molecules at a cost comparable to the evaluation of an interatomic potential
and much lower than the cost of evaluating a DFT model. When used correctly,
preconditioning leads to a substantial reduction in  the number of force calls
and thus is expected to significantly improve computing times
~\cite{1992RvMP...64.1045P,2016JChPh144p4109P}.

In this paper we introduce a simple yet effective way to precondition the
 standard NEB and string methods to obtain efficient and robust algorithms for
 computing MEPs in ill-conditioned geometries. Our scheme is further enhanced by
 a novel adaptive step length selection method to improve the robustness of the method.
 We demonstrate the effectiveness of this combination on a range of
 material modelling examples.

\section{The NEB and String methods}
Let $\bm{x}\in\mathbb{R}^M, M \in \mathbb{N},$ be a state, or configuration, of the dynamical
 system in question. We denote
 by $V(\bm{x})$ the potential
 energy of $\bm{x}$ and assume that
 $V$ is twice differentiable
 and that it has at least two local minima, which we denote by
 $\bm{x}_\mathrm{A}$ and $\bm{x}_\mathrm{B}$, separated by a single saddle point
 $\bm{x}_\mathrm{S}$ of Morse index 1 (to ensure that
 there is a unique direction of steepest descent at $\bm{x}_\mathrm{S}$
 \cite{0951-7715-24-6-008}). An MEP of
 the transition from $\bm{x}_\mathrm{A}$ to $\bm{x}_\mathrm{B}$ is defined as the
 intrinsically parametrised path $\bm{x}^*(s),\: s\in [0,1]$,
 satisfying
 \begin{equation}
   \nabla^\perp V(\bm{x}^*) \equiv \mathbf{0},
 \end{equation}
 with end points at the local minima
 $\bm{x}^*(0)=\bm{x}_\mathrm{A}$, $\bm{x}^*(1)=\bm{x}_\mathrm{B}$,
 where $\nabla^\perp V(\bm{x}) = \left(\mathbf{I} -
 \frac{\bm{x}'}{\|\bm{x}'\|}
 \otimes\frac{\bm{x}'}{\|\bm{x}'\|}\right)\nabla V(\bm{x})$ and where $\bm{x}' =
 \frac{\mathrm{d}\bm{x}}{\mathrm{d}s}$. (We note that, strictly speaking
 $\nabla^\perp V$ depends on $\bm{x}'$ as well as $\bm{x}$ but for the sake of
 simplicity of notation we will only write $\nabla^\perp V(\bm{x})$.) We only present our derivation of preconditioning and numerical tests for the original string method~\cite{PhysRevB.66.052301} but not the simplified string method~\cite{simplifiedString}, which seems to be used less in practise. However, this is not a fundamental restriction, and we expect no major changes when applying our preconditioning ideas to the simplified string method.

The NEB and string methods discretise a path $\bm{x}(s)$ by interpolating $N$
discrete points $\{\bm{x}_n\}_{n = 1}^N$. In the present work we will employ
cubic spline interpolation\cite{Dierckx:1993:CSF:151103}, imposing the
``not-a-knot'' boundary condition, but the methods we discuss can be readily
extended to other interpolation schemes as well.

To evolve the discrete path to equilibrium we introduce a pseudo-temporal
coordinate $\tau$ and write
$\dot{\bm{x}}=\frac{\mathrm{d}\bm{x}}{\mathrm{d}\tau}$.
 The evolution of $\bm{x}_n(\tau)$ is then described by the system of ODEs
 \begin{equation}\label{eqGradFlow}
 \dot{\bm{x}}_n =
 -\nabla^\perp V(\bm{x}_n) + \boldsymbol\eta,
 \end{equation}
 where $\bm{\eta} = 0$ leads to the string method, while the NEB method
 introduces elastic interactions between adjacent images along the path by
 adding the term
 $$
  \bm{\eta} = \bm{\eta}_{\rm neb} = \kappa
 \left(\bm{x}''\cdot\frac{\bm{x}'}{\|\bm{x}'\|}
 \right)\frac{\bm{x}'}{\|\bm{x}'\|}.
 $$

 The system \eqref{eqGradFlow} can be solved with any ordinary differential equation (ODE) numerical
 integrator. Most commonly, Euler's method~\cite{simplifiedString} is used,
 which yields an update step of the form
 \begin{equation}
 \label{eqStaticStringNEB}	
    \boxed{\bm{x}^{k+1}_n = \bm{x}^k_n+\alpha^k \left[-\nabla^\perp
    V(\bm{x}^k_n)+ \boldsymbol{\eta}_n^k\right],}
 \end{equation}
 where $\boldsymbol{\eta}_n^k=\boldsymbol{\eta}((\bm{x}^k_n)', (\bm{x}^k_n)'')$ and
 $\alpha^k$ is the timestep at iteration $k$.

While for NEB the presence of the elastic interaction $\bm{\eta}$ enforces an
approximate equidistribution of the nodes along the path, the string method
 reparametrises the path after each iteration to ensure that the
 images remain equidistant with respect to a suitable metric. In the
 continuous limit, as $N\rightarrow\infty$ a converged discretised
 path tends to the correct MEP, independently of the choice of
 the reparametrisation metric~\cite{Cameron2011}. We initially use the
 standard $\ell^2$-norm defined by $\|\bm{x}\|^2=\bm{x}\cdot\bm{x}$,
 but we will introduce a different notion of distance later on.

To summarise, the updating relations are given by \eqref{eqStaticStringNEB}
where, for the string method only, there is an additional redistribution of the
images after the update step. We follow precisely the approach described in
Eq.~12 in Ref.~\onlinecite{simplifiedString}, but for simplicity of presentation
do not make this step explicit.

The updating steps Eq.~\eqref{eqStaticStringNEB} for the string and NEB
 methods as well as the subsequent analysis were
 defined in terms of total derivatives of the path variable $\bm{x}$
 (i.e. in terms of $\bm{x}'$ and
 $\bm{x}''$), as they are motivated from the respective laws of
 classical dynamics. This information is available at each iteration at no extra
 cost as we use cubic spline interpolation to find an expression for $\bm{x}(s)$
 \cite{PhysRevB.66.052301,1998cqdc.conf..385J}.

\section{Preconditioning}
The NEB and string methods have slow convergence rates when they are subjected
 to ill-conditioned energy landscapes $V$.
 However, a suitable preconditioner
 $\mathbf{P}\in\mathbb{R}^{M\times M}$ that is cheap to compute can be
 used to reduce the condition number of the Hessian $\nabla\nabla V$
 along the path.
 In steepest descent optimisation, preconditioning has related but distinct
 interpretations: (a) as an
 approximation of the hessian, $\mathbf{P} \approx \nabla\nabla V$,
 in analogy to Newton's scheme or (b) as a
 coordinate transformation in the state space, $\bm{x} \mapsto \mathbf{P}^{1/2} \bm{x}$, that captures information of the
 local curvature of the potential landscape (mapping hyperellipsoids to balls)
 \cite{opacb1120179}.

We will now describe a preconditioning technique for NEB and string methods.
 The same preconditioners
 used in geometry optimisation of interatomic potentials~\cite{2016JChPh144p4109P,Mones2018}
 are expected to be valid for the purposes of preconditioning each image separately.
 We first present our construction of the preconditioned string method which has a simpler updating step.
\subsection{Preconditioned String Method}\label{secPreconString}
Let us first consider the simple case where $\mathbf{P}$ is constant in $\bm{x}$.
 Starting from the coordinate transformation
 \begin{equation}
    \bm{x}\mapsto\mathbf{P}^{-1/2}\bm{x}:=\tilde{\bm{x}},
 \end{equation}
 with corresponding $\tilde{V}(\tilde{\bm{x}}) =
 V(\mathbf{P}^{1/2} \tilde{\bm{x}})$,
 it is trivial to deduce that
 $\frac{\partial \tilde{x}_i}{\partial x_j} = P^{1/2}_{ij}$. The
 string method in the transformed space has updating step
 $\tilde{\bm{x}}_n^{k+1} = \tilde{\bm{x}}_n^k -
 \alpha^k \nabla^\perp \tilde{V}(\tilde{\bm{x}}_n^k)$
 which, for convenience we rewrite as
 \begin{alignat}{2}
    &\tilde{\bm{x}}_n^{k+1} &&= \tilde{\bm{x}}_n^k-\alpha^k \left(\mathbf{I} -
    \tilde{\bm{t}}_n^k \otimes \tilde{\bm{t}}_n^k
    \right)
    \nabla_{\tilde{x}} \tilde{V}(\tilde{\bm{x}}_n^k),\label{eqStringTransformed} \\
	 \notag
	 &\tilde{\bm{t}}_n^k &&= \frac{ (\tilde{\bm{x}}_n^k)' }{ \| (\tilde{\bm{x}}_n^k)' \|}.
 \end{alignat}
Reversing the coordinate transformation we obtain an equivalent formulation
in the original coordinates with updating step
 \begin{alignat}{2}
    \label{eqStringPreconConst}
    &\bm{x}^{k+1}_n &&= \bm{x}^k_n -\alpha^k \left(\mathbf{P}^{-1} -
    \bm{t}_{\mathrm{P},n}^k \otimes \bm{t}_{\mathrm{P},n}^k \right)
    \nabla_x V(\bm{x}_n^k), \\
	 \notag
	 &\bm{t}_{\mathrm{P},n}^k &&= \frac{(\bm{x}_n^k)'}{\|(\bm{x}_n^k)'\|_\mathbf{P}}.
 \end{alignat}
 where care needs to be taken to normalise the tangents
 $\bm{x}'$ with respect to the $\mathbf{P}$-norm,
 $\mathbf{\|\mathbf{y}\|_\mathbf{P}}=
 (\mathbf{y}\cdot\mathbf{P}\mathbf{y})^{1/2}$,
 instead of the usual $\ell^2$-norm,
 $\mathbf{\|\mathbf{y}\|}=(\mathbf{y}\cdot\mathbf{y})^{1/2}$.

Expressing the reparametrisation step in terms of coordinates in the
 configuration space is trivial, as it suffices to replace the usual
 $\ell^2$-norm with the $\mathbf{P}$-norm, due to linearity of the
 $\frac{\mathrm{d}}{\mathrm{d}s}$ operator.

The systems of interest, however, are described by preconditioners that are not
 constant in the configuration space~\cite{2016JChPh144p4109P}, which leads to
 a Riemannian metric framework and in particular the analogue of Eq.~\eqref{eqStringTransformed} involves the evaluation of
 $\nabla\mathbf{P}^{1/2}(\bm{x}_n^k)$ which is computationally
 expensive.
 We circumvent these issues entirely by dropping these terms. Preliminary tests (which we do not discuss here) showed that this does not lead to any loss of performance.
 Thus, we obtain the preconditioned string method
 \begin{equation}
    \label{eqStringPrecon}
    \bm{x}^{k+1}_n = \bm{x}^k_n- \alpha^k \nabla^\perp V_\mathrm{P}(\bm{x}_n^k),
 \end{equation}
 where we defined the quantity
 \begin{align*}
 	\nabla^\perp V_\mathrm{P}(\bm{x}_n^k) &= \left( [\mathbf{P}_n^k]^{-1} -
		\bm{t}_{\mathrm{P},n}^k \otimes \bm{t}_{\mathrm{P},n}^k \right)
 \nabla_x V(\bm{x}_n^k), \\
 \bm{t}_{\mathrm{P}, n}^k &= \frac{ (\bm{x}_n^k)'}{\|(\bm{x}_n^k)'\|_{\mathbf{P}^k_n}},
\end{align*}
 in terms of the $\mathbf{P}^k_n=\mathbf{P}(\bm{x}_n^k)$. We are left to specify how to re-parametrise the path.
 Recall that in the continuous limit, we are free to use any parametrisation for
 the path. In our setting, the premise is that $\|\cdot\|_{\mathbf{P}}$ is a
 more natural notion of distance than the standard $\ell^2$-norm $\|\cdot\|$, hence we will use the following notion of distance along the path:
 \begin{equation}\label{eqDistanceP}
    \mathrm{d}_{\mathbf{P}}(\bm{x}, \bm{y}) :=  \Bigg((\bm{x}-\mathbf{y})\cdot
    \bigg(\frac{\mathbf{P}(\bm{x})+
    \mathbf{P}(\mathbf{y})}{2}\bigg)
    (\bm{x}-\mathbf{y}) \Bigg)^{1/2}.
 \end{equation}
 We note that $\mathrm{d}_{\bf P}$ is {\it not} a metric in the technical sense, as it
 does not satisfy the triangle inequality. However, it is an approximation
 (discretisation) of the geodesic distance on the Riemannian manifold induced by
 the preconditioner ${\bf P}$, hence it is reasonable to expect that it can be
 used for the reparametrisation of the path. In practise, we have not
 encountered any difficulties related to this issue. The details of the
 preconditioned reparametrisation algorithm are given in Appendix~\ref{appReparametrisation}.

\subsection{Preconditioned NEB method}
An entirely analogous argument yields the preconditioned NEB method,
 \begin{equation}
    \bm{x}^{k+1}_n = \bm{x}^k_n +
    \alpha^k [-\nabla^\perp V_{\mathrm{P}}(\bm{x}_n^k)
    + (\bm{\eta}_{\mathrm{neb, P}})^k_n],
    \label{eqNEBPrecon}
 \end{equation}
where
$$
 (\bm{\eta}_{\mathrm{neb, P}})^k_n = \kappa 
 \left((\bm{x}_n^k)''\cdot\mathbf{P}_n^k
 \frac{(\bm{x}_n^k)'}{\|(\bm{x}_n^k)'\|_{\mathbf{P}_n^k}}
 \right)
 \frac{(\bm{x}_n^k)'}{\|(\bm{x}_n^k)'\|_{\mathbf{P}_n^k}}.
$$
\vspace{10pt}

Notice that this class of preconditioning schemes disregards the interactions
 between images and therefore, the preconditioner aids the convergence of the
 path only in the transverse direction. This is justified when the main source
 of ill-conditioning is due to the potential energy landscape, which is the case
 when only few images are used as is often done in practise.
 To summarise, the preconditioned updating relations are given by
 \begin{equation}
 \label{eqPreconStringNEB}
  \boxed{\bm{x}^{k+1}_n = \bm{x}^k_n+\alpha^k \left[-\nabla^\perp
     V_\mathrm{P}(\bm{x}_n^k) + (\bm{\eta}_\mathrm{P})_n^k\right],}
 \end{equation}
 where, in analogy to our earlier notation, $(\bm{\eta}_\mathrm{P})_n^k=0$ for the
 string method and $(\bm{\eta}_\mathrm{P})_n^k=(\bm{\eta}_{\mathrm{neb,P}})_n^k$
 for NEB.

\subsection{ODE solvers and steepest descent}

The optimisation step
 Eq.~\eqref{eqStaticStringNEB} was derived by applying Euler's
 method to the first order differential equation \eqref{eqGradFlow},
 but any ODE solver can be
 used instead. Here, we use an adaptive ODE solver based on
 Ref. \onlinecite{Hairer:1993:SOD:153158}
 to allow for some adaptivity in the step selection mechanism.

The user supplies an absolute and a relative tolerance $atol$ and $rtol$, which
 control the accuracy of the solution. We will demonstrate that choosing these
 two parameters is more intuitive and more robust than choosing the step length
 of the static method.

We modify an adaptive ODE solver, \textit{ode12}~\cite{Hairer:1993:SOD:153158}.
To begin we compute a trial
step ${\bm{x}}_n^{k+1}$ using Eq.~\eqref{eqPreconStringNEB} with a given step-length $\alpha^k$.
Next, we use ${\bm{x}}_n^{k+1}$ to compute a second-order solution to the underlying ODE system, via
\begin{alignat*}{2}
	\tilde{\bm{x}}_n^{k+1} = \bm{x}_n^k + {\textstyle \frac{1}{2}} \alpha^k
 	\big[ \mathbf{f}_n^k + \mathbf{f}_n^{k+1} \big],
\end{alignat*}
where $\mathbf{f}_n^k = - \nabla^\perp V_\mathrm{P}(\bm{x}_n^k) +
(\bm{\eta}_\mathrm{P})^k_n$ is the driving force on image $n$ at timestep $k$.
We can then use the difference $\tilde{\bm{x}}_n^{k+1} - {\bm{x}}_n^{k+1}$, or
equivalently the difference $\mathbf{f}_n^k - \mathbf{f}_n^{k+1}$ as an
error indicator.

Taking this as a starting point and following, for example,
Ref.~\onlinecite{odeTimestepping} to implement an adaptive time-stepping
algorithm we obtain an algorithm that underestimates the local error in the
neighbourhood of equilibria and in particular will not converge as $k \to
\infty$. To overcome this, we add a second step-length selection mechanism
based on minimising the residual. In essence, the adaptive ODE step selection
should be used in the pre-asymptotic regime while minimising the residual
is a suitable mechamism in the asymptotic regime.

This leads to the following step-length selection algorithm, which we label
\textit{ode12r}: we define the re-scaled residual error
\begin{equation}\label{eqResidualError}
	R^{k+1} =
	\max_n\left\| \mathbf{P}_n^k \nabla^\perp V_\mathrm{P}(\bm{x}_n^k) \right\|_\infty,
\end{equation}
and local error
 \begin{equation*}\label{eqlocalError}
 E^{k+1} =
 \max_{n,j}\left\{\frac{ \frac12 \big|(\mathbf{f}_n^k - \mathbf{f}_n^{k+1})_j \big|}
  {\max\left\{
  |(\bm{x}_n^k)_j|, |(\bm{x}_n^{k+1})_j|,
  \frac{atol}{rtol}\right\}}\right\}
\end{equation*}
where the index $j$ denotes vector components.
We then accept the proposed $\bm{x}_n^{k+1}$ if the scaled
residual error
 satisfies either one of the two following
 conditions:\vspace{2pt}\\
    \indent 1) $R^{k+1} \leq R^k(1-c_1\alpha^k)$,\vspace{2pt}\\
    \indent 2) $R^{k+1} \leq R^k c_2$ AND $E^{k+1} \leq rtol$,\vspace{2pt}\\
 for contraction and growth parameters $c_1$ and $c_2\in\mathbb{R}$.

 Whether the step is accepted or rejected, we now compute
   two step-length
  candidates using (1) the adaptive solver and (2) a simple line-search procedure.

  The step-length candidate given by the \textit{ode12} solver is
  $\alpha^{k+1}_{\rm ode12} = \frac12 \alpha^k \sqrt{\mathit{rtol}/E^{k+1}}$.
	For the second candidate,  we approximate the driving force along the previous search direction
  by its linear interpolant $(1-\theta) \mathbf{f}_n^k + \theta \mathbf{f}_n^{k+1}$. We then
  minimise $\| (1-\theta) \mathbf{f}_n^k + \theta \mathbf{f}_n^{k+1} \|_{\mathbf{P}_n^k}^2$ with respect to
  $\theta$ to obtain $\alpha^{k+1}_{\rm ls} =
  \theta \alpha^k$.

	If the current step $\bm{x}^{k+1}$ is accepted then the next step-length
	candidate is chosen to be
	\[
		\alpha^{k+1} = \max\big( {\textstyle \frac14} \alpha^k, \min\big(4\alpha^k, \alpha^{k+1}_{\rm ls}, \alpha^{k+1}_{\rm ode12}\big)\big).
		\]
   If the step $\bm{x}^{k+1}$ is rejected, then the new step-length candidate
	starting from $\bm{x}^k$ is
	\[
		\alpha^k = \max\big( {\textstyle \frac1{10}} \alpha^k, \min\big({\textstyle \frac14}\alpha^k, \alpha^{k+1}_{\rm ls}, \alpha^{k+1}_{\rm ode12}\big)\big).
	\]

\begin{figure}[ht]\centering
   \includegraphics[width=3.34in ,angle=0]{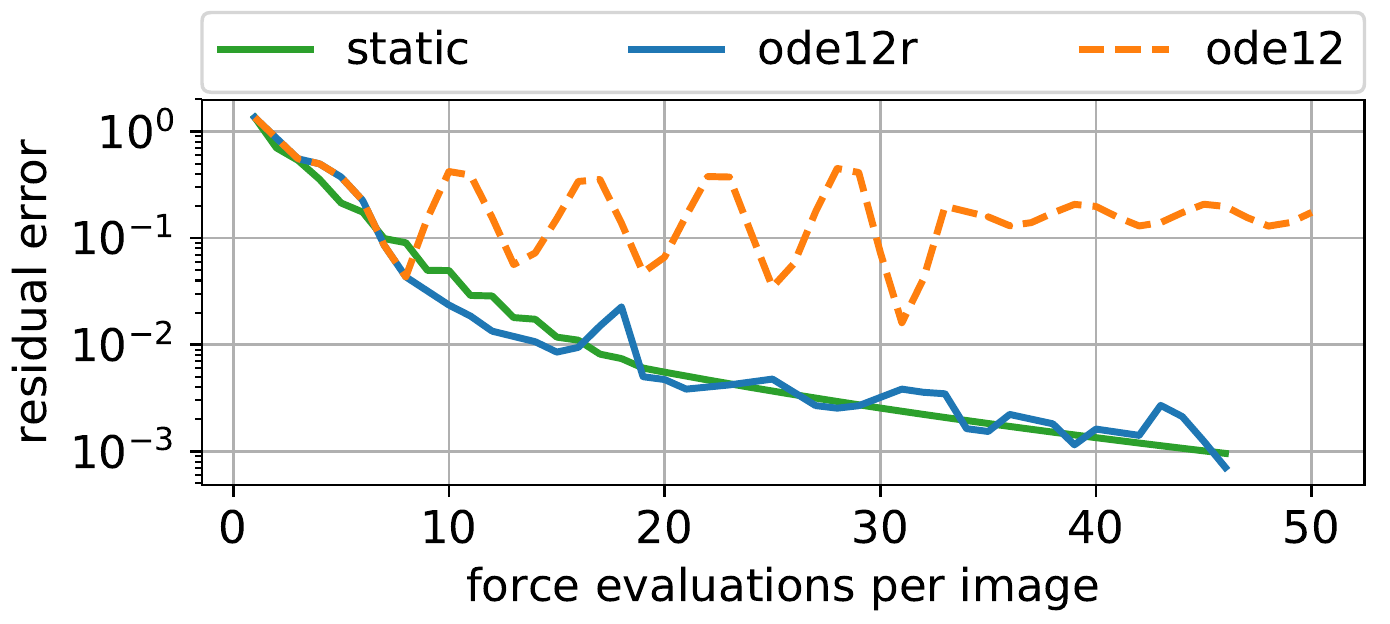}
   \caption{Convergence rate of the string method applied to vacancy
   migration in a 249-atom bcc W supercell modeled
   with the EAM4 potential~\cite{0953-8984-25-39-395502}. Optimal static
   time stepping, time stepping with \textit{ode12} and time stepping with \textit{ode12r}
   were used with a path consisting of 5 images.
   \label{figVac3Dodesolver}}
 \end{figure}

 Figure \ref{figVac3Dodesolver} demonstrates how \textit{ode12} effectively
 selects appropriate step lengths in the pre-asymptotic regime, but stagnates in
 the asymptotic regime for the case of vacancy migration in tungsten modelled
 with the EAM4 class of the Embedded Atom Model (EAM) interatomic potential
 proposed by Marinica \textit{et al.}~\cite{0953-8984-25-39-395502}. The
 convergence rate of the modified \textit{ode12r} agrees with the results of
 \textit{ode12} in the pre-asymptotic regime but successfully converges upon
 reaching the asymptotic regime.

\section{Results}\label{secResults}
We tested our preconditioning scheme for a variety of examples. First, we looked
 at examples using interatomic potentials which are not the main target, as
 these are typically fast models and constructing a preconditioner may not be
 computationally efficient relative to force evaluations. These examples,
 however, demonstrate how the number of force evaluations can be reduced with
 the use of the preconditioner. Further fine-tuning the preconditioner
 implementation and application (e.g., our current implementation updates the
 preconditioner after each iteration, which could be avoided) one would
 still obtain significant practical speed-ups for severely ill-conditioned
 cases.

 We then compare with a density functional
 theory (DFT) model to confirm our earlier results. In the following tables we
 compare the number of force evaluations per image needed to converge to
 `coarse' and `fine' target accuracies (maximum force less than
 $10^{-1}$~eV/\AA{} and $10^{-3}$~eV/\AA{}, respectively) using unpreconditioned
 and preconditioned schemes with either static or adaptive \textit{ode12r} step
 selection. The criterion for convergence is the magnitude of the residual error
 $R^{k+1}$ as defined in Eq.~\eqref{eqResidualError}. For the use of the
 \textit{ode12r} step selection, fitting the $rtol$ and $atol$ parameter was
 simple, as it was observed that $rtol=0.1$ was sufficient in most cases for
 convergence but other values $rtol=1$ and $rtol=0.01$ were occasionally more
 appropriate. The value of $atol$ was chosen so that $atol/rtol=1$ in all cases
 except the 2D vacancy of Sec.~\ref{secVac}, where $atol/rtol=0.01$ had to be
 used instead.

\subsection{Vacancy Migration}\label{secVac}

First we consider the diffusion of a vacancy in a two dimensional 60-atom
 triangular lattice governed by a Lennard-Jones potential $V(r) = 4\epsilon[(\sigma/r)^{12} - 2 (\sigma/r)^6]$ with parameters $\epsilon = 1.0$, $\sigma = 2^{-\frac{1}{6}}$.
 The vacancy is located at the centre of the cell initially and migrates in the
 $y$ direction by one lattice spacing. Periodic boundary conditions are imposed
 in the $x$ and $y$ directions. Table~\ref{tbl2DVacancy} shows
 the number of force calls per image required for convergence.
 The \textit{exponential preconditioner} (Exp) introduced in Packwood \textit{et
 al.}~\cite{2016JChPh144p4109P} with parameters $A=3.0$ and
 $r_\mathrm{cut}=2.5$, which utilises bond-connectivity information
 to treat the ill-conditioning of the system allowed convergence beyond the
 $10^{-3}$ tolerance, which the unpreconditioned case could not
 achieve within a reasonable number of iterations. The latter came as a surprise to
 us, as on the contrary to the real vacancy migration systems that we study
 next, this artificial set up exhibits more severe ill-conditioning. We note that for the unpreconditioned case when using the \textit{ode12r} time stepping for the string method we had to use $atol/rtol=0.01$. The absolute differences $\|\bm{x}_1-\bm{x}_2\|_\infty$ of the positions of any two converged paths at the images $\bm{x}_1$ and $\bm{x}_2$ nearest to the saddles, with and without preconditioning, were of the order of $8\times10^{-3}$.

 \begin{table}[ht]
   \centering
   \begin{tabular}{p{3.0cm} p{0.9cm} p{0.7cm} p{1.3cm} p{1.cm}}
   \multicolumn{5}{l}{2D Vacancy} \\
   \hline
   Step selection & \multicolumn{2}{c}{static} & \multicolumn{2}{l}{\textit{ode12r} solver}\\
   Tol  & $10^{-1}$ & $10^{-3}$ & $10^{-1}$ & $10^{-3}$\\
   \hline
   $\text{String}$ & $197$ & $*$ & $52$ & $*$\\
   $\text{String (p)}$ & $16$ & $38$ & $12$ & $33$\\
   $\text{NEB}$ & $200$ & $*$ & $53$ & $*$\\
   $\text{NEB (p)}$ & $19$ & $60$ & $14$ & $67$
   \end{tabular}
   \caption{Number of force evaluations per image required by the string and NEB methods
   to converge the vacancy migration MEP in a 9 image path of a 60-atom 2D cell
	 modelled with a Lennard-Jones potential,
   with either the static or \textit{ode12r} step length selection methods.
   In the cases marked *, the algorithm did not converge
	 within a reasonable number of iterations. \label{tbl2DVacancy}}
 \end{table}

Next, we considered a three dimensional system containing a vacancy, specifically
 a 107-atom Cu fcc supercell in a fixed cell with
 periodic boundary conditions. Interactions were modeled with a Morse potential
 with parameters $A = 4.0,\: \epsilon = 1.0$ and nearest neighbour distance
 $r_0 = 2.55$~\AA{}
 with interactions between atoms expressed by
 $V(r) = \epsilon( e^{-2A(r/r_0 - 1)} - 2e^{-A (r/r_0 - 1)})$.
 The exponential preconditioner introduced in Packwood
 \textit{et al.}~\cite{2016JChPh144p4109P} was used with parameters $A=3.0$ and
 $r_\mathrm{cut}=2.2r_0=5.62$~\AA{}.
 Table~\ref{tbl3DvacancyCu} shows the number of force evaluations per image
 needed for convergence to two preset tolerance limits.
 This example demonstrates how the \textit{ode12r} solver can aid the
 performance of the string and NEB methods if a static step is not suitable.
 Preconditioning gave almost a 2-fold speedup for the higher accuracy results, but no improvement for the lower acuracy. The absolute differences of the positions of the converged paths at the saddle, as done before, were well below $3\times10^{-14}$\AA{}.

 \begin{table}[ht]
  \centering
  \begin{tabular}{p{3.0cm} p{0.9cm} p{0.7cm} p{1.3cm} p{1.cm}}
   \multicolumn{5}{l}{Vacancy in Cu supercell} \\
   \hline
   Step selection & \multicolumn{2}{c}{static} & \multicolumn{2}{l}{\textit{ode12r} solver}\\
   Tol / eV/\AA{} & $10^{-1}$ & $10^{-3}$ & $10^{-1}$ & $10^{-3}$\\
     \hline
     $\text{String}$ & $8$ & $74$ & $8$ & $41$\\
     $\text{String (p)}$ & $7$ & $38$ & $8$ & $21$\\
     $\text{NEB}$ & $8$ & $57$ & $8$ & $27$\\
     $\text{NEB (p)}$ & $7$ & $37$ & $8$ & $19$
  \end{tabular}
  \caption{Force evaluations per image needed for the string and NEB methods
  for the migration of a vacancy in a $107$-atom
  Cu fcc supercell modelled by a Morse potential.
  The MEP was discretised with $5$ images.
  \label{tbl3DvacancyCu}
  }
 \end{table}

 A 53-atom W bcc
 supercell modelled with the EAM4 potential described in
 Ref. \onlinecite{0953-8984-25-39-395502} was examined as well.
 Periodic boundary conditions were imposed. A \textit{force field preconditioner} (FF) was constructed, by suitably modifying the EAM hessian to enforce positivity; see Mones \textit{et al.}~\cite[p. 9]{Mones2018} for full details. This yields
 up to 6 times faster convergence for higher accuracies as shown in
 Table~\ref{tbl3DvacancyW}. The absolute differences of the positions of the converged paths at the saddle, were well below $5\times10^{-9}$\AA{}.
 \begin{table}[ht]
 \centering
 \begin{tabular}{p{3.0cm} p{0.9cm} p{0.7cm} p{1.3cm} p{1.cm}}
   \multicolumn{5}{l}{Vacancy in W supercell} \\
   \hline
   Step selection & \multicolumn{2}{c}{static} & \multicolumn{2}{l}{ode12r solver}\\
   Tol / eV/\AA{} & $10^{-1}$ & $10^{-3}$ & $10^{-1}$ & $10^{-3}$\\
    \hline
    $\text{String}$ & $7$ & $77$ & $7$ & $49$\\
    $\text{String (p)}$ & $5$ & $12$ & $5$ & $9$\\
    $\text{NEB}$ & $8$ & $58$ & $7$ & $35$\\
    $\text{NEB (p)}$ & $5$ & $10$ & $8$ & $17$
 \end{tabular}
 \caption{Force evaluations per image needed for the string and NEB methods
 to converge the MEP for vacancy migration in a
 $53$-atom W bcc supercell modelled by the EAM4
 potential~\cite{0953-8984-25-39-395502}.
 The path was discretised by $5$ images and the preconditioner was
 constructed from the force field~\cite{Mones2018}.
 \label{tbl3DvacancyW}
 }
 \end{table}

We studied the same 53-atom W vacancy system with density functional theory
 (DFT), as implemented in the Castep~\cite{castep_paper} software.  The exchange
 correlation functional was approximated by the Perdew, Burke and Ernzerhof
 (PBE) generalised gradient approximation (GGA)\cite{PhysRevLett.77.3865}, with
 a planewave energy cut-off of $500$~eV and a $2\times2\times2$ Monkhorst-Pack
 grid to sample the Brillouin zone (a comparison of convergence behaviour
 obtained with a $3\times3\times3$ k-point grid was carried out which showed
 that the use of the $2\times2\times2$ k-point grid is sufficient). Step
 selection with \textit{ode12r} step and static step selection schemes was
 studied. A regularised FF preconditioner based on the EAM Hessian was used, $\mathbf{P}
 = (1-\lambda) \mathbf{P}_{\rm FF} + \lambda \mathbf{P}_{\rm Exp} + c \mathbf{I}$, where $c = 0.05$,
 $\lambda = 0.4$, $\mathbf{P}_{\rm FF}$ is described in Ref. \onlinecite[p. 9]{Mones2018},
 and the $\mathbf{P}_{\rm Exp}$ parameters were fitted to $\mathbf{P}_{\rm FF}$.


The path is made up of 5 images and traversing the path in subsequent iterations of the NEB
 and string methods was performed in an alternating order, allowing
 efficient reuse of previous electronic structure data to start the next optimisation step.

\begin{figure*}[ht]
	\begin{minipage}{.48\linewidth}
     \includegraphics[width=3.34in ,angle=0]{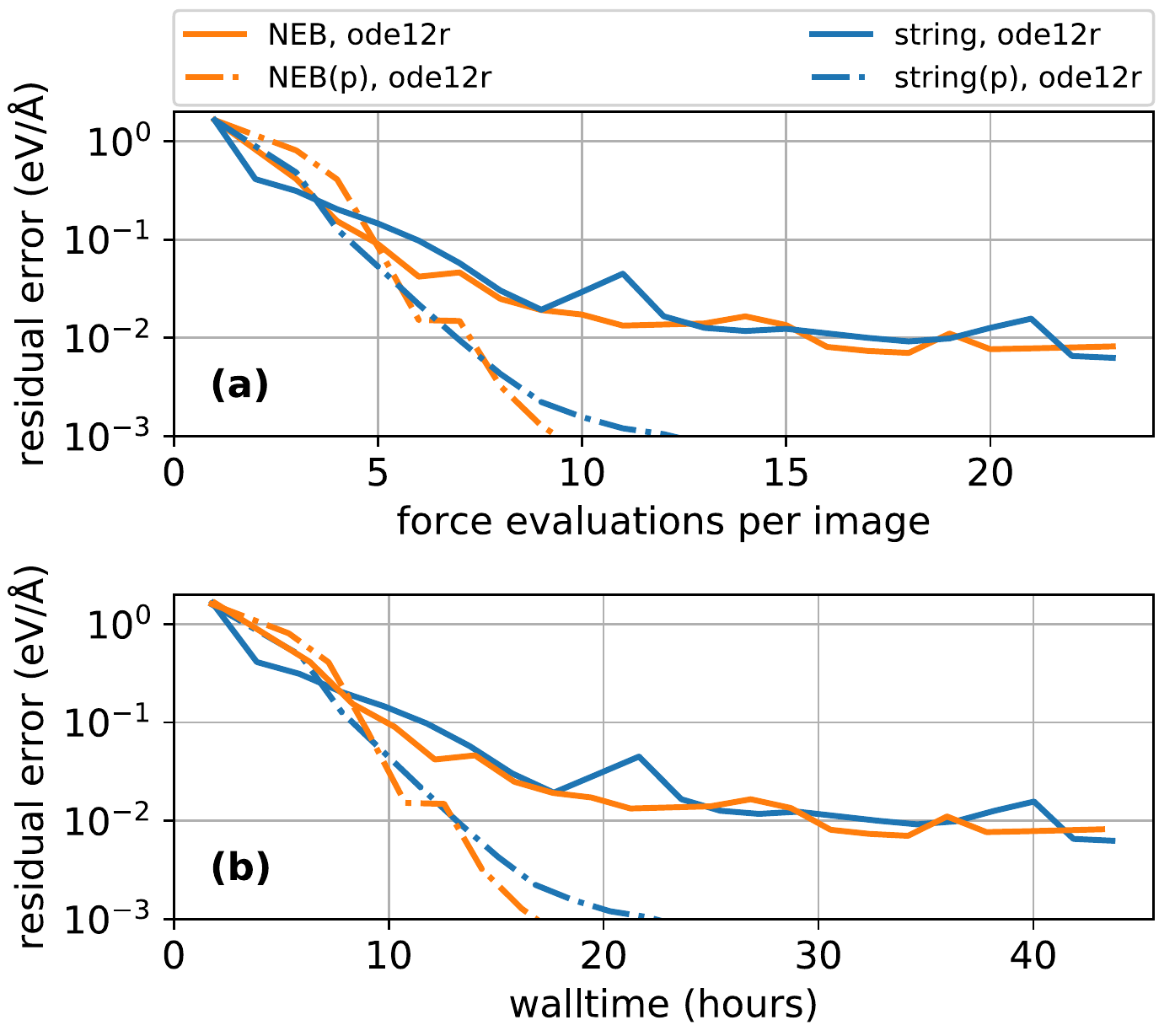}
\caption{Convergence of the string and NEB methods with and without
 preconditioner for a $53$-atom bcc W supercell containing a vacancy and modelled
 with DFT. The upper panel (a) shows the error as a function of the number of
 force evaluations per image and the lower (b) as a function of the time
 required to converge. Time stepping with \textit{ode12r} was used with a path of 5
 images. Comparison shows that constructing and evaluating the preconditioner
 is negligible compared to the cost of force computation.\label{figDFTode12vac}}
\end{minipage}
\hspace{\fill}
\begin{minipage}{.48\linewidth}
	 \includegraphics[width=3.34in ,angle=0]{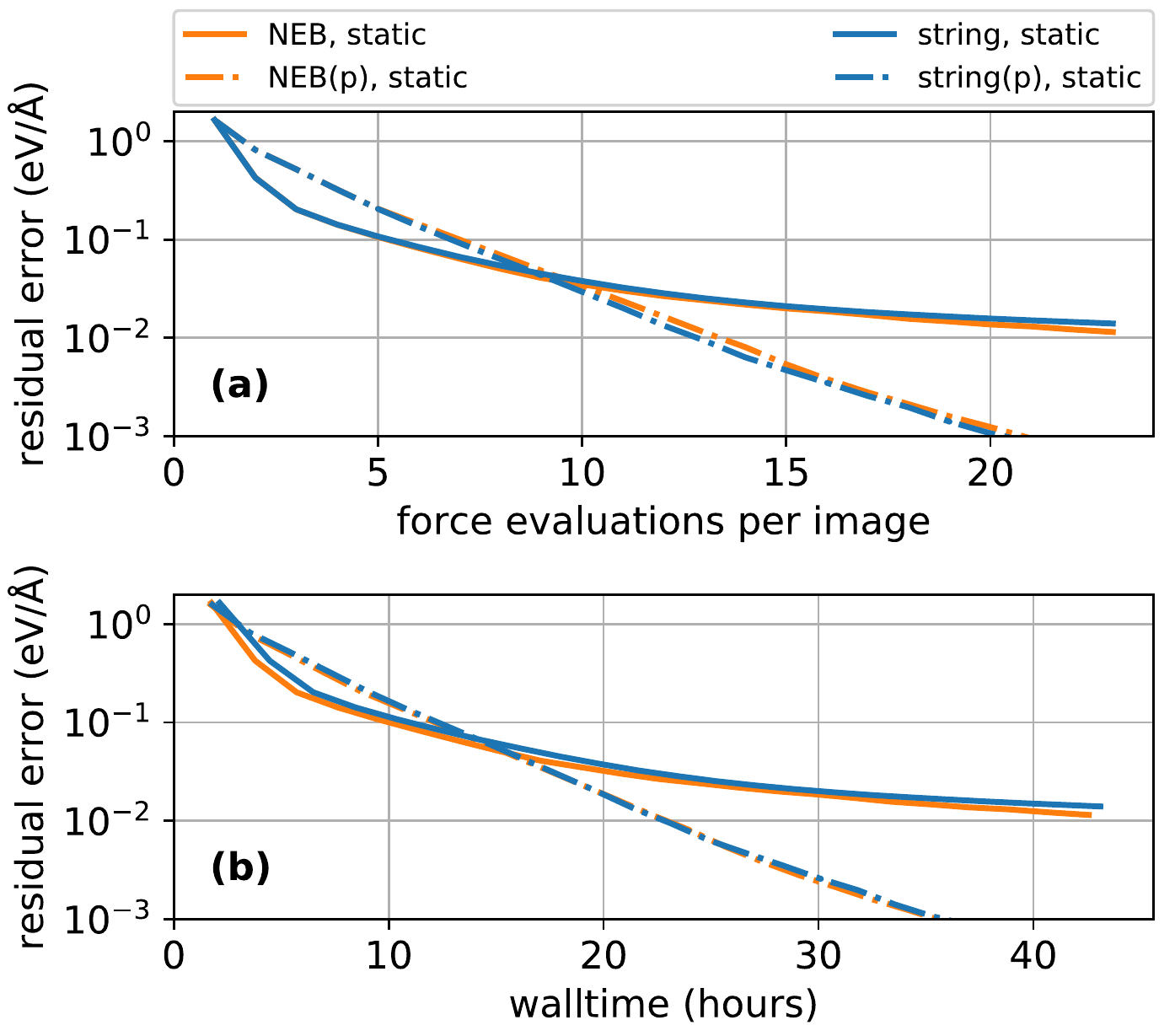}
\caption{Convergence of the string and NEB methods with and without
 preconditioner for a $53$-atom W bcc supercell containing a vacancy and modelled
 with DFT. The upper panel (a) shows the error as a function of the number of
 force evaluations per image and the lower panel (b) shows the error as a
 function of the time required to converge. The static time
 step was chosen by extrapolating the \textit{ode12r} data. The path was
 discretised by 5 images.\newline\label{figDFTstatvac}}\vspace{\fill}
\end{minipage}
\end{figure*}

Unlike the EAM case above, the
 preconditioner we used for the DFT model does not describe the potential energy
 surface of the DFT model exactly, but nevertheless gives a speed-up of a
 factor of two for an accuracy of $\sim 10^{-2}$~eV/\AA{} and furthermore
 allows accuracies of the order of
 $\sim 10^{-3}$~eV/\AA{} to be achieved, unlike the unpreconditioned case, as
 shown in Figs.~\ref{figDFTode12vac} and ~\ref{figDFTstatvac}. The results of Table~\ref{tbl3DvacancyW}
 suggest that constructing a better preconditioner would improve these results further.
 Notice further that the number of force evaluations needed for convergence and the time
 needed for convergence are in agreement (by comparison of the upper and
 lower panes of Figs.~\ref{figDFTode12vac} and \ref{figDFTstatvac}), confirming that the computational cost of
 constructing the preconditioner model is negligible compared to the cost of computing
 DFT forces, justifying our earlier assumptions. We note that the gain
 of preconditioning would be expected to further increase with system size~\cite{2016JChPh144p4109P}. The absolute differences of the positions of the converged paths at the saddle were of the order of $1\times10^{-4}$\AA{}.

\subsection{Screw Dislocation}

In the final example we study a $\frac{1}{2}\left\langle 111\right\rangle$
 screw dislocation in a 562-atom W bcc structure confined in a cylinder
 of radius equal to 20~{\AA} and surrounded by an 11~\AA{} cylindical shell of
 clamped atoms, with periodic boundary conditions along the dislocation line
 ($z$) direction. The
 system is simulated with the same EAM4 potential. The dislocation advances by one glide
 step. Table~\ref{tblScrewDislocation} shows the computational costs for
 converging the MEP with the NEB and string methods, using either static
 or \textit{ode12r} step length selection.
 A force field preconditioner built from the same EAM potential was used
 for geometry optimisation.

 \begin{table}[ht]
 \centering
 \begin{tabular}{p{3.0cm} p{0.9cm} p{0.7cm} p{1.3cm} p{1.cm}}
    \multicolumn{5}{l}{Screw Dislocation} \\
    \hline
    step selection & \multicolumn{2}{c}{static} & \multicolumn{2}{l}{ode12r solver}\\
    Tol / eV/\AA{} & $10^{-1}$ & $10^{-3}$ & $10^{-1}$ & $10^{-3}$\\
    \hline
    $\text{String}$ & $40$ & $272$ & $14$ & $124$\\
    $\text{String (p)}$ & $7$ & $48$ & $9$ & $21$\\
    $\text{NEB}$ & $40$ & $312$ & $14$ & $162$\\
    $\text{NEB (p)}$ & $7$ & $47$ & $7$ & $21$
 \end{tabular}
 \caption{Computational cost for the NEB and string methods for a screw
 dislocation in a 562-atom W bcc cylinder simulated with the EAM4 Marinica potential
 \cite{0953-8984-25-39-395502}.
 The circular boundary is fixed at a
 radius of $\mathrm{R}=20$\AA{}. Periodic boundary conditions were imposed
 in the $z$ direction. The path was discretised by $9$ points.
 \label{tblScrewDislocation}}
 \end{table}

Upon preconditioning, we observed a 5-fold speed up for the static case for low
accuracies but only a 2-fold speed up for the \textit{ode12r} case. For a
higher accuracy, a speed up of a factor of 6 was observed and
there was a speed up of a factor of at least 2 from using the
\textit{ode12r} step selection over the static step selection for both the
unpreconditioned and preconditioned cases. This indicates that the fitted static step is only suitable in the pre-asymptotic regime and a larger step size is suitable in the asymtotic regime, showcasing the advantages of using the adaptive \textit{ode12r} scheme over the hand-tuned static step. The absolute differences of the positions of the converged paths at the saddle were below $2\times10^{-3}$\AA{}.

\begin{figure}[ht]\centering
      \includegraphics[width=3.34in ,angle=0]{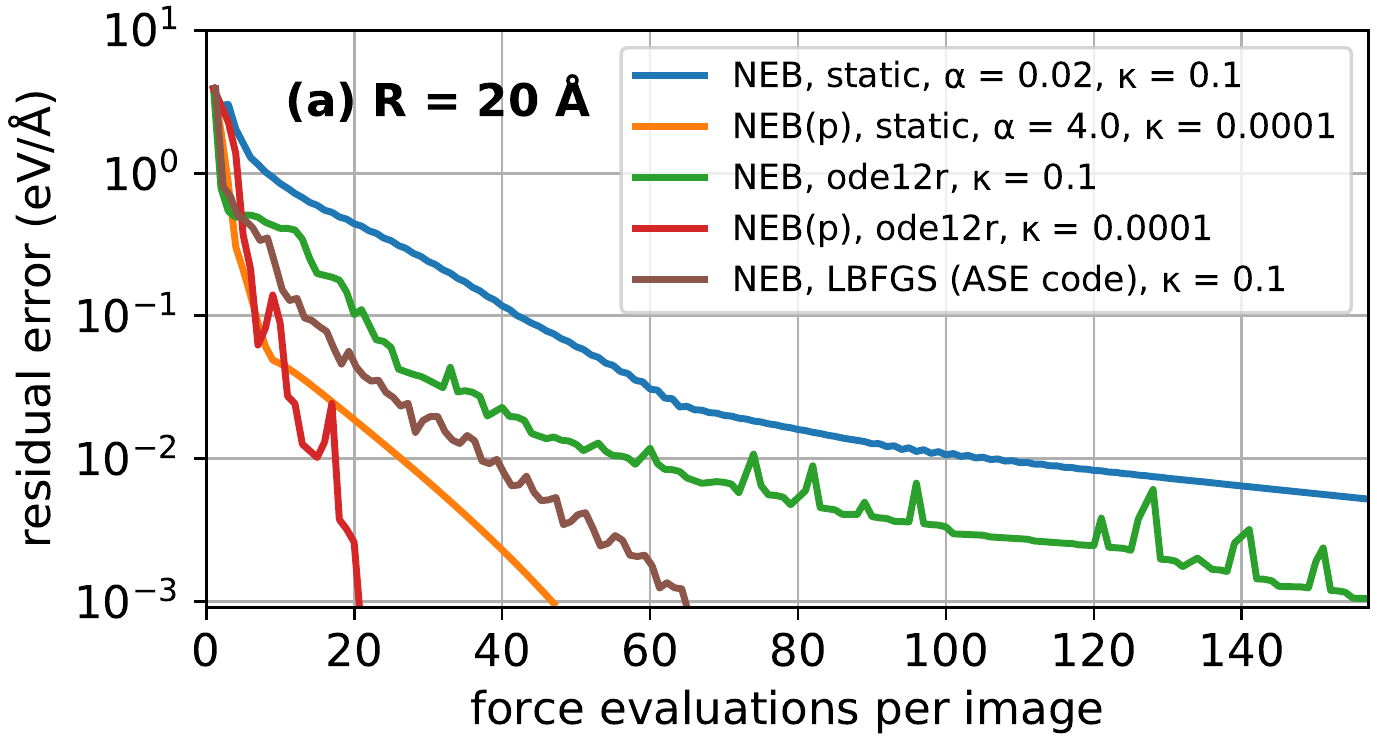}
      \includegraphics[width=3.34in ,angle=0]{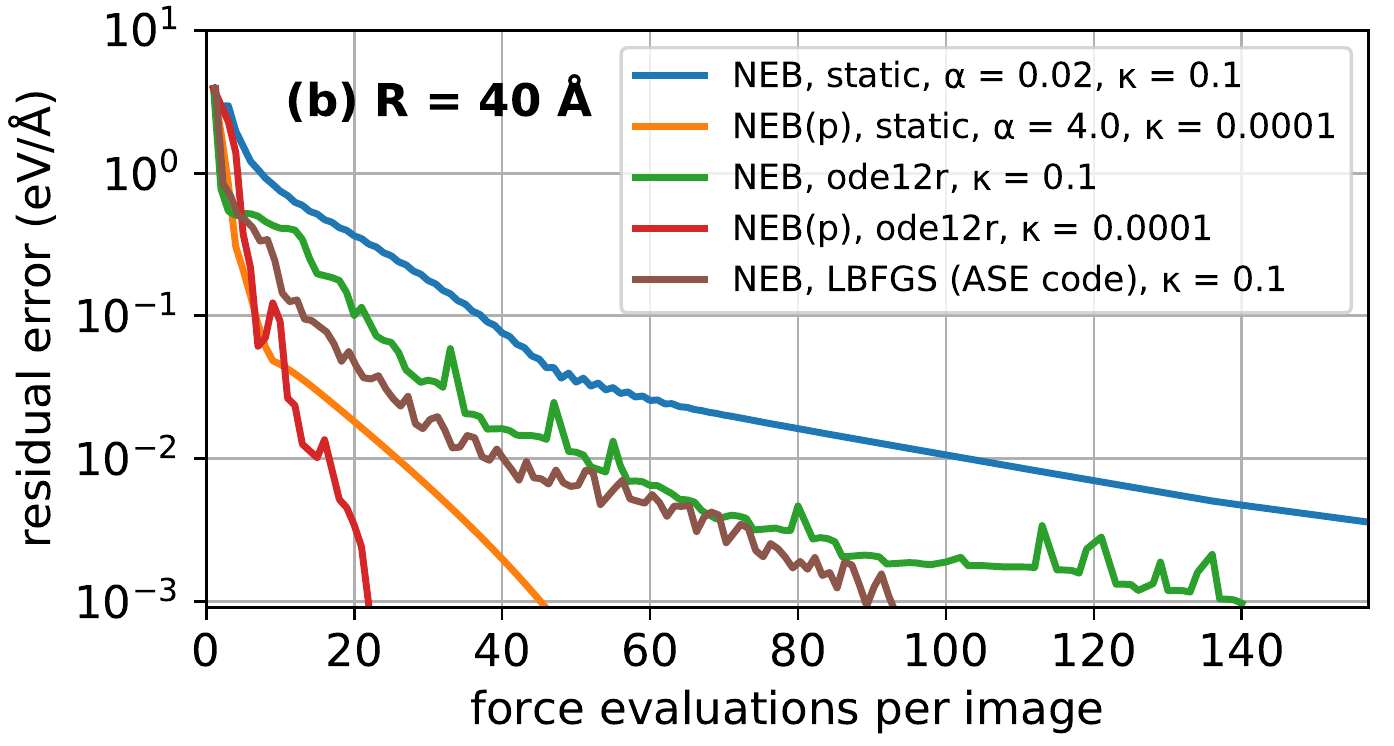}
 \caption{Convergence of NEB variants for a screw dislocation in a 562-atom W
 bcc cylindrical structure (a) and a 1489-atom W bcc cylindrical structure
 (b) modeled with the EAM4 Marinica
 potential~\cite{0953-8984-25-39-395502}. Atoms outside outer radii of
 $\mathrm{R}=20$\AA{} and $\mathrm{R}=40$\AA{} respectively were clamped, with
 periodic boundary conditions along the dislocation line. The path was
 discretised with 7 images (excluding the minima at each end, which were held
 fixed).\label{figDislocationNEB} The horizontal axis of the plots was cut after 160 force evaluations per image to focus on the performance of the preconditioned schemes. The static unpreconditioned NEB method converged after 312 force evaluations per image for the $\mathrm{R}=20$\AA{} case and after 343 force evaluations per image for the $\mathrm{R}=40$\AA{} case.}
 \end{figure}

We investigated this system further, focussing on the NEB implementation to
 allow comparison with the widely used Limited memory Broyden
 \hyp{} Fletcher \hyp{} Goldfarb \hyp{} Shanno (LBFGS)
 \cite{Liu1989} optimisation algorithm, which can be used with the NEB implementation
 \cite{doi:10.1063/1.1323224} in the Atomic Simulation Environment
 (ASE)\cite{Larsen2017}.
 This required fixing the endpoints of the
 path at the minima as is done in the ASE code. The comparison was carried out on
 systems of two sizes. A force field preconditioner
 was used as before for the preconditioned cases. Figure \ref{figDislocationNEB} shows the convergence rate of
 the various NEB schemes for a radius of 20{\AA} in the upper panel (a)
 and for a radius of 40{\AA} in the lower panel (b). Note that although LBFGS
 gave good
 convergence in the unpreconditioned case, it lacks robustness. This is because
 the force field of the NEB algorithm is
 not conservative, violating one of LBFGS's assumptions. LBFGS constructs a
 Hessian matrix corresponding to a scalar field, failing to capture the effects
 of the transport terms of the NEB force field. Moreover, the lack of the energy
 function prevents the use of line search, required to ensure the method's
 stability; in the ASE LBFGS implementation a
 heuristic is instead used to impose a maximum step length of 0.04~\AA{}. Furthermore, it should be noted that because our preconditioning scheme does not treat the longitudinal force components, it is inappropriate for us to use it together with the LBFGS method for MEP finding methods.

\section{CONCLUSIONS}
  We have demonstrated that MEP finding techniques such as the NEB and the
 string method can exhibit slow convergence rates due to poor search direction
 and step-length selection during the optimisation procedure.
 We have introduced a new optimisation technique combining an adaptive time-stepping scheme
 with preconditioning to address ill-conditioning of the
 energy landscape in directions transverse to the path and to allow
 faster convergence to the minimum energy path.

 We observed that our new scheme gives a significant speed up and improved
 robustness over currently used approaches for a range of systems using both
 force fields and DFT. Moreover, it allows higher accuracies to be reached than
 existing methods.

 However, our preconditioning scheme targets
 transverse ill-conditioning only. The longitudinal terms,
 (e.g. the NEB spring interactions) are unaffected by the preconditioner,
 suggesting that our scheme provides a baseline for further improvements.

 An open source prototype implementation of our technique is available at
 \url{https://github.com/cortner/SaddleSearch.jl}.

\begin{acknowledgments}

This work was supported by the Engineering and Physical Sciences Research
Council (EPSRC) under grants EP/P002188/1, EP/R012474/1, EP/J021377/1 and EP/R043612/1, by ERC
Starting Grant 335120, and by the Royal Society under grant number RG160691.
Computing facilities were provided by the Scientific Computing Research
Technology Platform of the University of Warwick with support from the Science
Research Investment Fund.
We thank Petr Grigorev for providing the screw dislocation configurations.

\end{acknowledgments}

\appendix
\section{Reparametrising in preconditioned string}\label{appReparametrisation}
The path reparametrisation described in Eq.~12 in
Ref.~\onlinecite{simplifiedString} assumes that the $\ell^2$-metric is used to
measure distance. Here, we briefly describe the modifications required when it
is replaced with the metric $\mathrm{d}_\mathbf{P}$ defined in \eqref{eqDistanceP}, used in the
preconditioned string method introduced in Sec.~\ref{secPreconString}.

After accepting an optimisation step $k$ of Eq.~\eqref{eqStringPrecon} the
following steps are performed:
 \begin{enumerate}
   \item Compute the relative distances $\mathrm{d}_{\mathbf{P}}(\bm{x}_n^k, \bm{x}_{n-1}^k)$
	between the images $\{\bm{x}_n^k\}_n$, for all $n=2,\dots,N$.
	\item Define
   \begin{align}
     &s_1 = 0,\\\notag
     &s_n = \frac{\sum_{m=2}^{n}{\mathrm{d}_{\mathbf{P}}(\bm{x}_m^k, \bm{x}_{m-1}^k)}}
     {\sum_{m=2}^{N}{\mathrm{d}_{\mathbf{P}}(\bm{x}_m^k, \bm{x}_{m-1}^k)}},\: \mathrm{for}\:
     n=2,\dots,M.
   \end{align}
   \item Use cubic spline interpolation~\cite{Dierckx:1993:CSF:151103} of
    $\{s_n, \bm{x}_n^k\}_{n=1}^N$ to obtain
   $\bm{x}^k\left(s\right): [0,1]\rightarrow \mathbb{R}^N$.
   \item The new images are then given by
   \begin{equation}
     \bm{x}_{n}^k = \bm{x}^k\big({\textstyle \frac{n-1}{N-1}}\big),
	  \qquad	n = 1, \dots, N.
   \end{equation}
 \end{enumerate}

This algorithm does {\it not} ensure that images will be equidistributed
according to $\mathrm{d}_\mathbf{P}$. However it does ensure that images remain bounded away from
one another, which is the key property required for the string method.


\begin{thebibliography}{24}%
\makeatletter
\providecommand \@ifxundefined [1]{%
 \@ifx{#1\undefined}
}%
\providecommand \@ifnum [1]{%
 \ifnum #1\expandafter \@firstoftwo
 \else \expandafter \@secondoftwo
 \fi
}%
\providecommand \@ifx [1]{%
 \ifx #1\expandafter \@firstoftwo
 \else \expandafter \@secondoftwo
 \fi
}%
\providecommand \natexlab [1]{#1}%
\providecommand \enquote  [1]{``#1''}%
\providecommand \bibnamefont  [1]{#1}%
\providecommand \bibfnamefont [1]{#1}%
\providecommand \citenamefont [1]{#1}%
\providecommand \href@noop [0]{\@secondoftwo}%
\providecommand \href [0]{\begingroup \@sanitize@url \@href}%
\providecommand \@href[1]{\@@startlink{#1}\@@href}%
\providecommand \@@href[1]{\endgroup#1\@@endlink}%
\providecommand \@sanitize@url [0]{\catcode `\\12\catcode `\$12\catcode
  `\&12\catcode `\#12\catcode `\^12\catcode `\_12\catcode `\%12\relax}%
\providecommand \@@startlink[1]{}%
\providecommand \@@endlink[0]{}%
\providecommand \url  [0]{\begingroup\@sanitize@url \@url }%
\providecommand \@url [1]{\endgroup\@href {#1}{\urlprefix }}%
\providecommand \urlprefix  [0]{URL }%
\providecommand \Eprint [0]{\href }%
\providecommand \doibase [0]{http://dx.doi.org/}%
\providecommand \selectlanguage [0]{\@gobble}%
\providecommand \bibinfo  [0]{\@secondoftwo}%
\providecommand \bibfield  [0]{\@secondoftwo}%
\providecommand \translation [1]{[#1]}%
\providecommand \BibitemOpen [0]{}%
\providecommand \bibitemStop [0]{}%
\providecommand \bibitemNoStop [0]{.\EOS\space}%
\providecommand \EOS [0]{\spacefactor3000\relax}%
\providecommand \BibitemShut  [1]{\csname bibitem#1\endcsname}%
\let\auto@bib@innerbib\@empty
\bibitem [{\citenamefont {Voter}\ \emph {et~al.}(2002)\citenamefont {Voter},
  \citenamefont {Montalenti},\ and\ \citenamefont {Germann}}]{Voter2002}%
  \BibitemOpen
  \bibfield  {author} {\bibinfo {author} {\bibfnamefont {A.~F.}\ \bibnamefont
  {Voter}}, \bibinfo {author} {\bibfnamefont {F.}~\bibnamefont {Montalenti}}, \
  and\ \bibinfo {author} {\bibfnamefont {T.~C.}\ \bibnamefont {Germann}},\
  }\href@noop {} {\bibfield  {journal} {\bibinfo  {journal} {Annual Review of
  Materials Research}\ }\textbf {\bibinfo {volume} {32}},\ \bibinfo {pages}
  {321} (\bibinfo {year} {2002})}\BibitemShut {NoStop}%
\bibitem [{\citenamefont {{Eyring}}(1935)}]{1935JChPh...3..107E}%
  \BibitemOpen
  \bibfield  {author} {\bibinfo {author} {\bibfnamefont {H.}~\bibnamefont
  {{Eyring}}},\ }\href@noop {} {\bibfield  {journal} {\bibinfo  {journal}
  {Journal of Computational Physics}\ }\textbf {\bibinfo {volume} {3}},\
  \bibinfo {pages} {107} (\bibinfo {year} {1935})}\BibitemShut {NoStop}%
\bibitem [{\citenamefont {{Pollak}}\ and\ \citenamefont
  {{Talkner}}(2005)}]{reactionRateTheoryReview}%
  \BibitemOpen
  \bibfield  {author} {\bibinfo {author} {\bibfnamefont {E.}~\bibnamefont
  {{Pollak}}}\ and\ \bibinfo {author} {\bibfnamefont {P.}~\bibnamefont
  {{Talkner}}},\ }\href@noop {} {\bibfield  {journal} {\bibinfo  {journal}
  {Chaos}\ }\textbf {\bibinfo {volume} {15}} (\bibinfo {year}
  {2005})}\BibitemShut {NoStop}%
\bibitem [{\citenamefont {Vineyard}(1957)}]{VINEYARD1957121}%
  \BibitemOpen
  \bibfield  {author} {\bibinfo {author} {\bibfnamefont {G.~H.}\ \bibnamefont
  {Vineyard}},\ }\href@noop {} {\bibfield  {journal} {\bibinfo  {journal}
  {Journal of Physics and Chemistry of Solids}\ }\textbf {\bibinfo {volume}
  {3}},\ \bibinfo {pages} {121} (\bibinfo {year} {1957})}\BibitemShut {NoStop}%
\bibitem [{\citenamefont {Voter}(2007)}]{10.1007/978-1-4020-5295-8_1}%
  \BibitemOpen
  \bibfield  {author} {\bibinfo {author} {\bibfnamefont {A.~F.}\ \bibnamefont
  {Voter}},\ }\href@noop {} {\bibfield  {journal} {\bibinfo  {journal}
  {Radiation Effects in Solids. NATO Science Series}\ }\textbf {\bibinfo
  {volume} {235}},\ \bibinfo {pages} {1} (\bibinfo {year} {2007})}\BibitemShut
  {NoStop}%
\bibitem [{\citenamefont {{E}}\ \emph {et~al.}(2002)\citenamefont {{E}},
  \citenamefont {{Ren}},\ and\ \citenamefont
  {{Vanden-Eijnden}}}]{PhysRevB.66.052301}%
  \BibitemOpen
  \bibfield  {author} {\bibinfo {author} {\bibfnamefont {W.}~\bibnamefont
  {{E}}}, \bibinfo {author} {\bibfnamefont {W.}~\bibnamefont {{Ren}}}, \ and\
  \bibinfo {author} {\bibfnamefont {E.}~\bibnamefont {{Vanden-Eijnden}}},\
  }\href@noop {} {\bibfield  {journal} {\bibinfo  {journal} {Phys. Rev. B}\
  }\textbf {\bibinfo {volume} {66}},\ \bibinfo {pages} {052301} (\bibinfo
  {year} {2002})}\BibitemShut {NoStop}%
\bibitem [{\citenamefont {{E}}\ \emph {et~al.}(2007)\citenamefont {{E}},
  \citenamefont {{Ren}},\ and\ \citenamefont
  {{Vanden-Eijnden}}}]{simplifiedString}%
  \BibitemOpen
  \bibfield  {author} {\bibinfo {author} {\bibfnamefont {W.}~\bibnamefont
  {{E}}}, \bibinfo {author} {\bibfnamefont {W.}~\bibnamefont {{Ren}}}, \ and\
  \bibinfo {author} {\bibfnamefont {E.}~\bibnamefont {{Vanden-Eijnden}}},\
  }\href@noop {} {\bibfield  {journal} {\bibinfo  {journal} {The Journal of
  Chemical Physics}\ }\textbf {\bibinfo {volume} {126}},\ \bibinfo {eid}
  {164103} (\bibinfo {year} {2007})}\BibitemShut {NoStop}%
\bibitem [{\citenamefont {{Cameron}}\ \emph {et~al.}(2011)\citenamefont
  {{Cameron}}, \citenamefont {{Kohn}},\ and\ \citenamefont
  {{Vanden-Eijnden}}}]{Cameron2011}%
  \BibitemOpen
  \bibfield  {author} {\bibinfo {author} {\bibfnamefont {M.}~\bibnamefont
  {{Cameron}}}, \bibinfo {author} {\bibfnamefont {R.~V.}\ \bibnamefont
  {{Kohn}}}, \ and\ \bibinfo {author} {\bibfnamefont {E.}~\bibnamefont
  {{Vanden-Eijnden}}},\ }\href@noop {} {\bibfield  {journal} {\bibinfo
  {journal} {Journal of Nonlinear Science}\ }\textbf {\bibinfo {volume} {21}},\
  \bibinfo {pages} {193} (\bibinfo {year} {2011})}\BibitemShut {NoStop}%
\bibitem [{\citenamefont {{J{\'o}nsson}}\ \emph {et~al.}(1998)\citenamefont
  {{J{\'o}nsson}}, \citenamefont {{Mills}},\ and\ \citenamefont
  {{Jacobsen}}}]{1998cqdc.conf..385J}%
  \BibitemOpen
  \bibfield  {author} {\bibinfo {author} {\bibfnamefont {H.}~\bibnamefont
  {{J{\'o}nsson}}}, \bibinfo {author} {\bibfnamefont {G.}~\bibnamefont
  {{Mills}}}, \ and\ \bibinfo {author} {\bibfnamefont {K.~W.}\ \bibnamefont
  {{Jacobsen}}},\ }\href@noop {} {\bibfield  {journal} {\bibinfo  {journal}
  {Classical and Quantum Dynamics in Condensed Phase Simulations}\ ,\ \bibinfo
  {pages} {385}} (\bibinfo {year} {1998})}\BibitemShut {NoStop}%
\bibitem [{\citenamefont {{Henkelman}}\ and\ \citenamefont
  {{Jónsson}}(2000)}]{doi:10.1063/1.1323224}%
  \BibitemOpen
  \bibfield  {author} {\bibinfo {author} {\bibfnamefont {G.}~\bibnamefont
  {{Henkelman}}}\ and\ \bibinfo {author} {\bibfnamefont {H.}~\bibnamefont
  {{Jónsson}}},\ }\href@noop {} {\bibfield  {journal} {\bibinfo  {journal}
  {The Journal of Chemical Physics}\ }\textbf {\bibinfo {volume} {113}},\
  \bibinfo {pages} {9978} (\bibinfo {year} {2000})}\BibitemShut {NoStop}%
\bibitem [{\citenamefont {{Nocedal}}\ and\ \citenamefont
  {{Wright}}(2006)}]{opacb1120179}%
  \BibitemOpen
  \bibfield  {author} {\bibinfo {author} {\bibfnamefont {J.}~\bibnamefont
  {{Nocedal}}}\ and\ \bibinfo {author} {\bibfnamefont {S.~J.}\ \bibnamefont
  {{Wright}}},\ }\href@noop {} {}Springer Series in Operations Research and
  Financial Engineering\ (\bibinfo  {publisher} {Springer},\ \bibinfo {address}
  {Berlin},\ \bibinfo {year} {2006})\BibitemShut {NoStop}%
\bibitem [{\citenamefont {{Packwood}}\ \emph {et~al.}(2016)\citenamefont
  {{Packwood}}, \citenamefont {{Kermode}}, \citenamefont {{Mones}},
  \citenamefont {{Bernstein}}, \citenamefont {{Woolley}}, \citenamefont
  {{Gould}}, \citenamefont {{Ortner}},\ and\ \citenamefont
  {{Cs{\'a}nyi}}}]{2016JChPh144p4109P}%
  \BibitemOpen
  \bibfield  {author} {\bibinfo {author} {\bibfnamefont {D.}~\bibnamefont
  {{Packwood}}}, \bibinfo {author} {\bibfnamefont {J.}~\bibnamefont
  {{Kermode}}}, \bibinfo {author} {\bibfnamefont {L.}~\bibnamefont {{Mones}}},
  \bibinfo {author} {\bibfnamefont {N.}~\bibnamefont {{Bernstein}}}, \bibinfo
  {author} {\bibfnamefont {J.}~\bibnamefont {{Woolley}}}, \bibinfo {author}
  {\bibfnamefont {N.}~\bibnamefont {{Gould}}}, \bibinfo {author} {\bibfnamefont
  {C.}~\bibnamefont {{Ortner}}}, \ and\ \bibinfo {author} {\bibfnamefont
  {G.}~\bibnamefont {{Cs{\'a}nyi}}},\ }\href@noop {} {\bibfield  {journal}
  {\bibinfo  {journal} {The Journal of Chemical Physics}\ }\textbf {\bibinfo
  {volume} {144}},\ \bibinfo {pages} {164109} (\bibinfo {year}
  {2016})}\BibitemShut {NoStop}%
\bibitem [{\citenamefont {Mones}\ \emph {et~al.}(2018)\citenamefont {Mones},
  \citenamefont {Ortner},\ and\ \citenamefont {Csanyi}}]{Mones2018}%
  \BibitemOpen
  \bibfield  {author} {\bibinfo {author} {\bibfnamefont {L.}~\bibnamefont
  {Mones}}, \bibinfo {author} {\bibfnamefont {C.}~\bibnamefont {Ortner}}, \
  and\ \bibinfo {author} {\bibfnamefont {G.}~\bibnamefont {Csanyi}},\
  }\href@noop {} {\bibfield  {journal} {\bibinfo  {journal} {Scientific
  Reports}\ }\textbf {\bibinfo {volume} {8}},\ \bibinfo {pages} {13991}
  (\bibinfo {year} {2018})}\BibitemShut {NoStop}%
\bibitem [{\citenamefont {Lindh}\ \emph {et~al.}(1995)\citenamefont {Lindh},
  \citenamefont {Bernhardsson}, \citenamefont {Karlstr{\"o}m},\ and\
  \citenamefont {Malmqvist}}]{LINDH1995423}%
  \BibitemOpen
  \bibfield  {author} {\bibinfo {author} {\bibfnamefont {R.}~\bibnamefont
  {Lindh}}, \bibinfo {author} {\bibfnamefont {A.}~\bibnamefont {Bernhardsson}},
  \bibinfo {author} {\bibfnamefont {G.}~\bibnamefont {Karlstr{\"o}m}}, \ and\
  \bibinfo {author} {\bibfnamefont {P.-{\AA}.}\ \bibnamefont {Malmqvist}},\
  }\href@noop {} {\bibfield  {journal} {\bibinfo  {journal} {Chemical Physics
  Letters}\ }\textbf {\bibinfo {volume} {241}},\ \bibinfo {pages} {423 }
  (\bibinfo {year} {1995})}\BibitemShut {NoStop}%
\bibitem [{\citenamefont {{Payne}}\ \emph {et~al.}(1992)\citenamefont
  {{Payne}}, \citenamefont {{Teter}}, \citenamefont {{Allan}}, \citenamefont
  {{Arias}},\ and\ \citenamefont {{Joannopoulos}}}]{1992RvMP...64.1045P}%
  \BibitemOpen
  \bibfield  {author} {\bibinfo {author} {\bibfnamefont {M.~C.}\ \bibnamefont
  {{Payne}}}, \bibinfo {author} {\bibfnamefont {M.~P.}\ \bibnamefont
  {{Teter}}}, \bibinfo {author} {\bibfnamefont {D.~C.}\ \bibnamefont
  {{Allan}}}, \bibinfo {author} {\bibfnamefont {T.~A.}\ \bibnamefont
  {{Arias}}}, \ and\ \bibinfo {author} {\bibfnamefont {J.~D.}\ \bibnamefont
  {{Joannopoulos}}},\ }\href@noop {} {\bibfield  {journal} {\bibinfo  {journal}
  {Reviews of Modern Physics}\ }\textbf {\bibinfo {volume} {64}},\ \bibinfo
  {pages} {1045} (\bibinfo {year} {1992})}\BibitemShut {NoStop}%
\bibitem [{\citenamefont {{E}}\ and\ \citenamefont
  {{Zhou}}(2011)}]{0951-7715-24-6-008}%
  \BibitemOpen
  \bibfield  {author} {\bibinfo {author} {\bibfnamefont {W.}~\bibnamefont
  {{E}}}\ and\ \bibinfo {author} {\bibfnamefont {X.}~\bibnamefont {{Zhou}}},\
  }\href@noop {} {\bibfield  {journal} {\bibinfo  {journal} {Nonlinearity}\
  }\textbf {\bibinfo {volume} {24}},\ \bibinfo {pages} {1831} (\bibinfo {year}
  {2011})}\BibitemShut {NoStop}%
\bibitem [{\citenamefont {{Dierckx}}(1993)}]{Dierckx:1993:CSF:151103}%
  \BibitemOpen
  \bibfield  {author} {\bibinfo {author} {\bibfnamefont {P.}~\bibnamefont
  {{Dierckx}}},\ }\href@noop {} {}\ (\bibinfo  {publisher} {Oxford University
  Press, Inc.},\ \bibinfo {address} {New York, NY, USA},\ \bibinfo {year}
  {1993})\BibitemShut {NoStop}%
\bibitem [{\citenamefont {{Hairer}}\ \emph {et~al.}(1993)\citenamefont
  {{Hairer}}, \citenamefont {{N{\o}rsett}},\ and\ \citenamefont
  {{Wanner}}}]{Hairer:1993:SOD:153158}%
  \BibitemOpen
  \bibfield  {author} {\bibinfo {author} {\bibfnamefont {E.}~\bibnamefont
  {{Hairer}}}, \bibinfo {author} {\bibfnamefont {S.~P.}\ \bibnamefont
  {{N{\o}rsett}}}, \ and\ \bibinfo {author} {\bibfnamefont {G.}~\bibnamefont
  {{Wanner}}},\ }\href@noop {} {}\ (\bibinfo  {publisher} {Springer-Verlag New
  York, Inc.},\ \bibinfo {address} {New York, NY, USA},\ \bibinfo {year}
  {1993})\BibitemShut {NoStop}%
\bibitem [{\citenamefont {{Lamba}}(2000)}]{odeTimestepping}%
  \BibitemOpen
  \bibfield  {author} {\bibinfo {author} {\bibfnamefont {H.}~\bibnamefont
  {{Lamba}}},\ }\href@noop {} {\bibfield  {journal} {\bibinfo  {journal} {BIT
  Numerical Mathematics}\ }\textbf {\bibinfo {volume} {40}},\ \bibinfo {pages}
  {314} (\bibinfo {year} {2000})}\BibitemShut {NoStop}%
\bibitem [{\citenamefont {{Marinica}}\ \emph {et~al.}(2013)\citenamefont
  {{Marinica}}, \citenamefont {{Ventelon}}, \citenamefont {{Gilbert}},
  \citenamefont {{Proville}}, \citenamefont {S~L~{Dudarev}}, \citenamefont
  {{Marian}}, \citenamefont {{Bencteux}},\ and\ \citenamefont
  {{Willaime}}}]{0953-8984-25-39-395502}%
  \BibitemOpen
  \bibfield  {author} {\bibinfo {author} {\bibfnamefont {M.-C.}\ \bibnamefont
  {{Marinica}}}, \bibinfo {author} {\bibfnamefont {L.}~\bibnamefont
  {{Ventelon}}}, \bibinfo {author} {\bibfnamefont {M.~R.}\ \bibnamefont
  {{Gilbert}}}, \bibinfo {author} {\bibfnamefont {L.}~\bibnamefont
  {{Proville}}}, \bibinfo {author} {\bibfnamefont {S.~L.}\ \bibnamefont
  {S~L~{Dudarev}}}, \bibinfo {author} {\bibfnamefont {J.}~\bibnamefont
  {{Marian}}}, \bibinfo {author} {\bibfnamefont {G.}~\bibnamefont
  {{Bencteux}}}, \ and\ \bibinfo {author} {\bibfnamefont {F.}~\bibnamefont
  {{Willaime}}},\ }\href@noop {} {\bibfield  {journal} {\bibinfo  {journal}
  {Journal of Physics: Condensed Matter}\ }\textbf {\bibinfo {volume} {25}},\
  \bibinfo {pages} {395502} (\bibinfo {year} {2013})}\BibitemShut {NoStop}%
\bibitem [{\citenamefont {{Clark}}\ \emph {et~al.}(2005)\citenamefont
  {{Clark}}, \citenamefont {{Segall}}, \citenamefont {{Pickard}}, \citenamefont
  {{Hasnip}}, \citenamefont {{Probert}}, \citenamefont {{Refson}},\ and\
  \citenamefont {{Payne}}}]{castep_paper}%
  \BibitemOpen
  \bibfield  {author} {\bibinfo {author} {\bibfnamefont {S.}~\bibnamefont
  {{Clark}}}, \bibinfo {author} {\bibfnamefont {M.}~\bibnamefont {{Segall}}},
  \bibinfo {author} {\bibfnamefont {C.}~\bibnamefont {{Pickard}}}, \bibinfo
  {author} {\bibfnamefont {P.}~\bibnamefont {{Hasnip}}}, \bibinfo {author}
  {\bibfnamefont {M.}~\bibnamefont {{Probert}}}, \bibinfo {author}
  {\bibfnamefont {K.}~\bibnamefont {{Refson}}}, \ and\ \bibinfo {author}
  {\bibfnamefont {M.}~\bibnamefont {{Payne}}},\ }\href@noop {} {\bibfield
  {journal} {\bibinfo  {journal} {Zeitschrift für Kristallographie -
  Crystalline Materials}\ }\textbf {\bibinfo {volume} {220}},\ \bibinfo {pages}
  {567} (\bibinfo {year} {2005})}\BibitemShut {NoStop}%
\bibitem [{\citenamefont {Perdew}\ \emph {et~al.}(1996)\citenamefont {Perdew},
  \citenamefont {Burke},\ and\ \citenamefont
  {Ernzerhof}}]{PhysRevLett.77.3865}%
  \BibitemOpen
  \bibfield  {author} {\bibinfo {author} {\bibfnamefont {J.~P.}\ \bibnamefont
  {Perdew}}, \bibinfo {author} {\bibfnamefont {K.}~\bibnamefont {Burke}}, \
  and\ \bibinfo {author} {\bibfnamefont {M.}~\bibnamefont {Ernzerhof}},\
  }\href@noop {} {\bibfield  {journal} {\bibinfo  {journal} {Phys. Rev. Lett.}\
  }\textbf {\bibinfo {volume} {77}},\ \bibinfo {pages} {3865} (\bibinfo {year}
  {1996})}\BibitemShut {NoStop}%
\bibitem [{\citenamefont {{Liu}}\ and\ \citenamefont
  {{Nocedal}}(1989)}]{Liu1989}%
  \BibitemOpen
  \bibfield  {author} {\bibinfo {author} {\bibfnamefont {D.~C.}\ \bibnamefont
  {{Liu}}}\ and\ \bibinfo {author} {\bibfnamefont {J.}~\bibnamefont
  {{Nocedal}}},\ }\href@noop {} {\bibfield  {journal} {\bibinfo  {journal}
  {Mathematical Programming}\ }\textbf {\bibinfo {volume} {45}},\ \bibinfo
  {pages} {503} (\bibinfo {year} {1989})}\BibitemShut {NoStop}%
\bibitem [{\citenamefont {Larsen}\ \emph {et~al.}(2017)\citenamefont {Larsen},
  \citenamefont {Mortensen}, \citenamefont {Blomqvist}, \citenamefont
  {Castelli}, \citenamefont {Christensen}, \citenamefont {Du{\l}ak},
  \citenamefont {Friis}, \citenamefont {Groves}, \citenamefont {Hammer},
  \citenamefont {Hargus}, \citenamefont {Hermes}, \citenamefont {Jennings},
  \citenamefont {Jensen}, \citenamefont {Kermode}, \citenamefont {Kitchin},
  \citenamefont {Kolsbjerg}, \citenamefont {Kubal}, \citenamefont {Kaasbjerg},
  \citenamefont {Lysgaard}, \citenamefont {Maronsson}, \citenamefont {Maxson},
  \citenamefont {Olsen}, \citenamefont {Pastewka}, \citenamefont {Peterson},
  \citenamefont {Rostgaard}, \citenamefont {Schi{\o}tz}, \citenamefont
  {Sch{\"u}tt}, \citenamefont {Strange}, \citenamefont {Thygesen},
  \citenamefont {Vegge}, \citenamefont {Vilhelmsen}, \citenamefont {Walter},
  \citenamefont {Zeng},\ and\ \citenamefont {Jacobsen}}]{Larsen2017}%
  \BibitemOpen
  \bibfield  {author} {\bibinfo {author} {\bibfnamefont {A.~H.}\ \bibnamefont
  {Larsen}}, \bibinfo {author} {\bibfnamefont {J.~J.}\ \bibnamefont
  {Mortensen}}, \bibinfo {author} {\bibfnamefont {J.}~\bibnamefont
  {Blomqvist}}, \bibinfo {author} {\bibfnamefont {I.~E.}\ \bibnamefont
  {Castelli}}, \bibinfo {author} {\bibfnamefont {R.}~\bibnamefont
  {Christensen}}, \bibinfo {author} {\bibfnamefont {M.}~\bibnamefont
  {Du{\l}ak}}, \bibinfo {author} {\bibfnamefont {J.}~\bibnamefont {Friis}},
  \bibinfo {author} {\bibfnamefont {M.~N.}\ \bibnamefont {Groves}}, \bibinfo
  {author} {\bibfnamefont {B.}~\bibnamefont {Hammer}}, \bibinfo {author}
  {\bibfnamefont {C.}~\bibnamefont {Hargus}}, \bibinfo {author} {\bibfnamefont
  {E.~D.}\ \bibnamefont {Hermes}}, \bibinfo {author} {\bibfnamefont {P.~C.}\
  \bibnamefont {Jennings}}, \bibinfo {author} {\bibfnamefont {P.~B.}\
  \bibnamefont {Jensen}}, \bibinfo {author} {\bibfnamefont {J.}~\bibnamefont
  {Kermode}}, \bibinfo {author} {\bibfnamefont {J.~R.}\ \bibnamefont
  {Kitchin}}, \bibinfo {author} {\bibfnamefont {E.~L.}\ \bibnamefont
  {Kolsbjerg}}, \bibinfo {author} {\bibfnamefont {J.}~\bibnamefont {Kubal}},
  \bibinfo {author} {\bibfnamefont {K.}~\bibnamefont {Kaasbjerg}}, \bibinfo
  {author} {\bibfnamefont {S.}~\bibnamefont {Lysgaard}}, \bibinfo {author}
  {\bibfnamefont {J.~B.}\ \bibnamefont {Maronsson}}, \bibinfo {author}
  {\bibfnamefont {T.}~\bibnamefont {Maxson}}, \bibinfo {author} {\bibfnamefont
  {T.}~\bibnamefont {Olsen}}, \bibinfo {author} {\bibfnamefont
  {L.}~\bibnamefont {Pastewka}}, \bibinfo {author} {\bibfnamefont
  {A.}~\bibnamefont {Peterson}}, \bibinfo {author} {\bibfnamefont
  {C.}~\bibnamefont {Rostgaard}}, \bibinfo {author} {\bibfnamefont
  {J.}~\bibnamefont {Schi{\o}tz}}, \bibinfo {author} {\bibfnamefont
  {O.}~\bibnamefont {Sch{\"u}tt}}, \bibinfo {author} {\bibfnamefont
  {M.}~\bibnamefont {Strange}}, \bibinfo {author} {\bibfnamefont {K.~S.}\
  \bibnamefont {Thygesen}}, \bibinfo {author} {\bibfnamefont {T.}~\bibnamefont
  {Vegge}}, \bibinfo {author} {\bibfnamefont {L.}~\bibnamefont {Vilhelmsen}},
  \bibinfo {author} {\bibfnamefont {M.}~\bibnamefont {Walter}}, \bibinfo
  {author} {\bibfnamefont {Z.}~\bibnamefont {Zeng}}, \ and\ \bibinfo {author}
  {\bibfnamefont {K.~W.}\ \bibnamefont {Jacobsen}},\ }\href@noop {} {\bibfield
  {journal} {\bibinfo  {journal} {J. Phys. Condens. Matter}\ }\textbf {\bibinfo
  {volume} {29}},\ \bibinfo {pages} {273002} (\bibinfo {year}
  {2017})}\BibitemShut {NoStop}%
\end{thebibliography}
\end{document}